\documentclass[prb,twocolumn,aps]{revtex4}
\usepackage{footnote}
\usepackage{longtable,setspace}
\usepackage{graphicx,rotating,lscape}
\usepackage{color}
\usepackage{float,epsfig,amsmath,amssymb,euscript,color}
\usepackage{amsmath,amssymb,epsfig,capt-of,ifthen,calc}
\usepackage{float,euscript}
\usepackage{aecompl}
\usepackage{color}
\usepackage{soul}
\usepackage{multirow}
\usepackage{dcolumn}
\usepackage{longtable}
\usepackage{supertabular}
\usepackage{setspace}
\usepackage{ulem,enumitem}
\usepackage{bm}
\usepackage{hyperref}
\usepackage{natbib}

\usepackage{amsthm}

\definecolor{bleuclair}{rgb}{0.7, 0.7, 1.0}
\definecolor{rosepale}{rgb}{1.0, 0.7, 1.0}

\begin{document}

\title{Structural properties of water confined by phospholipid membranes}

\author{Fausto Martelli$^{1}$, Hsin-Yu Ko$^{1}$, Carles Calero Borallo$^{2,3}$, Giancarlo Franzese$^{2,3}$}
\affiliation{%
$^{1}$ Department of Chemistry, Princeton University, Princeton New
  Jersey, 08544 USA \\
  $^{2}$     Secci\'o de
    F\'isica Estad\'istica i Interdisciplin\`aria--Departament de
    F\'isica de la Mat\`eria Condensada, Facultat de F\'isica
    Universitat de Barcelona,  
    Martí i Franquès 1, 08028, Barcelona, Spain. \\
    $^{3}$ 
    Institute of Nanoscience and Nanotechnology (IN2UB), Universitat de Barcelona
    Av. Joan XXIII S/N, 08028 Barcelona,
    Spain}
%    \\
%faustom@princeton.edu, hsinyu@princeton.edu, ccalero@ffn.ub.es, gfranzese@ub.edu\\
%$^{3}$Center for Polymer Studies and Department of Physics, Boston
  %University, 590 Commonwealth Avenue, Boston, MA 02215, USA;}

\email{faustom@princeton.edu, gfranzese@ub.edu}

\begin{abstract}
Biological membranes are essential for the cell life and hydration
water provides the driving force for their assembly and stability.
Here we study the structural properties of water in a phospholipid membrane.
We characterize local 
structures inspecting the intermediate range order (IRO) adopting a
sensitive local order metric, recently proposed by Martelli et al., 
which measures and grades the 
degree of overlap of local environments with structures of perfect ice. 
Close to the membrane, water acquires high  IRO and 
changes its dynamical properties, e.g., slowing down 
its translational and rotational degrees of freedom in a region that
extends over $\simeq 1$ nm from the membrane interface. Surprisingly,
we show that at a distance as far as
$\simeq 2.5$ nm from the interface,
although the bulk-like dynamics is recovered,
water's IRO is still slightly higher than in bulk at 
the same thermodynamic conditions.
Therefore, the water-membrane interface has a structural effect at
ambient conditions that
propagates further than the often-invoked $1$ nm-length scale, a results
that should be taken carefully into account when analyzing
experimental data of water confined by membranes and could help us
understanding the role of water in biological systems.
\end{abstract}

\maketitle

\section{Introduction}\label{introduction}
Biological membranes provide a limiting structure which separates the 
interior and exterior of cells and organelles. Being selectively permeable, 
they control the flow of substances in and out of the cell, which permits to 
regulate the composition of the cell and the communication between cells 
through signalling. Membranes are also involved in the capture and release of energy. \par
Biological membranes are composed of a variety of biomolecules, including 
proteins, sugars, cholesterol, and phospholipids. Among those components, 
phospholipids provide the structure to biological membranes, since they can 
spontaneously self-assemble due to the hydrophobic effect into bilayers. 
For this reason, phospholipid membranes are used as a model to investigate 
basic properties of biological membranes both in experimental and theoretical 
studies. Among a wide variety of phospholipids, dimyristoylphosphatidylcholine 
(DMPC) are phospholipids with a choline as a headgroup and a tailgroup formed 
by two myristoyl chains. Choline based phospholipids are ubiquitous in cell 
membranes and used in drug targeting liposomes~\cite{Hamley}.\par
Hydration water is fundamental for the
proper functioning of biological membranes. Indeed, it is known that the
presence of water strongly influences the stability, fluidity and phase
behavior of phospholipid membranes, which determines their
function. In addition, hydration water mediates the interaction of
biological membranes with other biomolecules and with ions. Due to its
importance, hydration water at phospholipid membranes has been
extensively studied experimentally~\cite{Fitter_JPhysChemB1999,
Trapp_JCP2010,Wassall_BiophysJ1996,Righini_PRL2007,Zhao_Fayer_JACS2008, Tielrooij_BiophysJ2009,Hua_CPC2015} 
and using computer simulations~\cite{Rog_ChemPhysLett2002,Bhide_JCP2005,Berkowitz_chemrev2006,Hansen_PRL2013,Zhang_Berkowitz_JPhysChemB2009,Gruenbaum_JChemPhys_2011,Calero_Mat2016}.
The results show that water is abundant in the interfacial region of bilayers (lipid
headgroups), and can even penetrate into deeper regions of the
membrane~\cite{Bhide_JCP2005,Calero_Mat2016}. As a result, fully
hydrated DMPC
can take $\approx$30 hydration
water molecules per lipid. Furthermore, computer simulations show that
water molecules form strong hydrogen bonds with lipid phosphate
groups, as well as with the carbonyl oxygens of phosphatidylcholine
lipids~\cite{Rog_ChemPhysLett2002, Bhide_JCP2005}, slowing down
both water orientational~\cite{Righini_PRL2007, Zhao_Fayer_JACS2008, Tielrooij_BiophysJ2009,
Zhang_Berkowitz_JPhysChemB2009, Gruenbaum_JChemPhys_2011,Calero_Mat2016} 
and translational~\cite{Wassall_BiophysJ1996,Calero_Mat2016} dynamics.
At low hydration water is kept inside the membrane revealing its
essential role to guarantee structural stability~\cite{Calero_Mat2016}.
In highly hydrated systems, i.e., when lipid surfaces are in contact with many water molecules,
dynamical properties of bulk water such as diffusivity and rotational dynamics, and density 
are recovered at relatively short distances from the lipid surface (see, e.g., Calero \textit{et al.}~\cite{Calero_Mat2016}). 
On the other hand, structural characterization of water in such systems are lacking, mostly because structural 
properties are counted for grant once the density is known.\par
In this article, we characterize the local structure 
at the intermediate range order (IRO),
i.e. the second shell of neighbors,
of water molecules confined in  between two membranes, each made of a lipid bilayer.
We employ a 
sensitive local order metric (LOM) recently introduced by Martelli et al.%
~\cite{Martelli_2016}. The LOM characterizes the molecular order in the neighborhood of a
site in a liquid, an amorphous or a crystal.
We compare the IRO of each water molecule with the IRO of perfect
cubic, I$c$, ice and we grade the degree 
of deviation from the perfect I$c$ structure~\footnote{Similar results hold when using the second shell of neighbors 
on hexagonal ice as a reference structure}. 
We measure the extent to which the lipid surfaces affect the structure
of water at different distances from the surface, and we relate structural changes with dynamical changes. 
We find that the membrane perturbs the IRO of water at a distance of at least $\simeq 2.5$ nm, while bulk dynamical 
properties are recovered closer to the membrane ($\sim 1.5$ nm), a result that be should takene carefully into account 
when analyzing experimental and/or theoretical results of confined water in general, and that could provide further insights 
in understanding water properties under such conditions~\cite{gallo_2017}.
%, therefore extending the 
%previously believed range of perturbation (citation here).
Water molecules acquire more and more 
local order moving to the proximity of the fluctuating lipid surfaces. 
This effect is small but
not negligible, and is accompanied by a slowing down in the translational and rotational degrees of freedom,
in agreement with previous experimental and numerical observations~\cite{Righini_PRL2007, Zhao_Fayer_JACS2008, Tielrooij_BiophysJ2009,Zhang_Berkowitz_JPhysChemB2009, Gruenbaum_JChemPhys_2011,Calero_Mat2016,Wassall_BiophysJ1996}.
Interestingly, approaching the interfaces water molecules show a sudden drop in the translational degrees 
of freedom, to values comparable to water at deeply undercooled conditions.
We also observe that, approaching the lipid interfaces, the rotations around the water dipole are less affected 
then the rotations round the $\overrightarrow{\rm{OH}}$ vector, indicating that the interactions between P--groups and water hydrogens
are stronger--and more probable--than the interactions between N-heads and water oxygen.\\
The article is organized as follows. In Section~\ref{methods} we present the details of the LOM and 
the details of the numerical simulations. In Section~\ref{results} we show 
our findings. Our conclusions and final remarks are presented in Section~\ref{conclusions}

\section{Methods}\label{methods}
\subsection{The local order metric}
Recently Martelli et al.~\cite{Martelli_2016}
introduced a LOM
that measures the degree of order in the neighborhood of an atomic or molecular site in a condensed 
medium. \par
The local environment of a water molecule $i$ ($i=1, \dots, N$) in a configuration
defines a local {\it pattern} formed by $M$ neighboring sites.
Here we 
include only the oxygen atoms second neighbors of the oxygen site $i$.
%There are $N$ local patterns, one for each atomic site $i$ in the
%system.
Indicating by $\mathbf{P}^{i}_{j}$ ($j=1,\dots, M$) the coordinates 
in the laboratory frame of the $M$ neighbors of pattern $i$, we define
the centroid as 
$\mathbf{P}_{c}^{i}\equiv\frac{1}{M}\sum_{j=1}^{M}\mathbf{P}^{i}_{j}$
and the rescaled positions as
%In the following we refer the positions of the sites of the pattern to their centroid, i.e.
$\mathbf{\widetilde{P}}^{i}_{j} \equiv \mathbf{P}^{i}_{j}-\mathbf{P}_{c}^{i}$.
%\rightarrow \mathbf{P}_{i}^{j}$.

To compare the local order of the pattern
$i$ with a {\it reference}  structure, we
choose $M$ sites in a given lattice and set the lattice
step equal to the $i$th O--O {\it bond unit} (BU) 
$d_i\equiv \frac{1}{M}\sum_{j=1}^{M}|\mathbf{\widetilde{P}}^{i}_{j}|$,
i.e. the average equilibrium $i$-$j$ distance in the pattern.
The pattern-$i$ centroid and the 
reference-$i$ centroid are set to coincide, while
the reference orientation is arbitrary. 
%The sites of the pattern and of the reference are labeled by the indices $i$ of the position vectors. 
The local order metric $S(i)$ at site $i$ is the maximum of the
overlap function
when the reference orientation is varied 
and the pattern
indices are permuted, i.e.
\begin{equation}
  S(i)\equiv
  \max_{\theta,\phi,\psi;\mathcal{P}}\prod_{j=1}^{M}\exp\left(-\frac{\left|
    \mathbf{P}_{j_{\mathcal{P}}}^{i}-\mathbf{R}_{j}^{i}\right|^2}{2\sigma^{2}M}\right), 
  \label{eq:Eq1}
\end{equation}
where $\theta,\phi,\psi$ are Euler angles for a given orientation of the reference, 
$i_{\mathcal{P}}$ are the permuted indices of the pattern sites corresponding to a permutation $\mathcal{P}$,
and $\sigma\equiv d/4.4$ is a parameter that controls the spread of the Gaussian functions. This value is chosen 
such that the tails of the Gaussians spread to half of the interatomic distance between oxygens in the second 
shell of neighbors in perfect cubic ice at ambient conditions.
As a consequence of the point symmetry of the reference the overlap function defined in Eq.~(\ref{eq:Eq1})
has multiple equivalent maxima. 
To compute $S(i)$
%(Eq.~\ref{eq:Eq1})
it is sufficient to locate only one of these maxima.
The 
point symmetry of the reference allows us
to explore
only a fraction $1/L$ of the Euler angle domain $\Omega$, 
which we call $\Omega/L$ , the irreducible domain of the Euler angles, being $L$ the 
number of proper point symmetry operations of the reference. Inside $\Omega/L$ we pick 
at random with uniform probability $15$ orientations and we select permutations by assigning nearest 
pattern-reference pairs. With the best alignement out of the $15$ orientations, we fix the 
permutation and optimize the orientation via conjugate gradient. \par
The LOM is an intrinsic property of the local environment at variance
with the straightforward overlap
function
%$\mathcal{O}(j)$
that would depend on the orientation of the reference and on the ordering of the
sites in the pattern.
The LOM satisfies the inequalities 
$0 \lesssim S(i) \leq 1$. The two limits correspond, respectively, to a completely disordered local 
pattern ($S(i)\rightarrow 0$) and to an ordered local pattern matching perfectly the reference ($S(i)\rightarrow 1$),
therefore grading each
local environment on an increasing scale of local order from zero to one. \par
We define an order parameters based on $S(i)$ as the average score $S$, or site
averaged LOM,
\begin{equation}
  S\equiv \frac{1}{N}\sum_{i=1}^{N}S(i)
  \label{eq:Eq3}
\end{equation}
and we refer to $S$ as \textit{score}.

\subsection{Simulations details}\label{simulations}

We perform molecular dynamics (MD) simulations
of $7040$ water molecules confined between two layers of 
$128$ DMPC lipids~\footnote{We use here a higher 
number of water molecules with respect to that considered in
Ref.~\cite{Calero_Mat2016} for a similar  system}.
We use periodic boundary conditions in all directions in such a way
that our systems corresponds to water confined between the two
different sides of the same lipid bilayer. The bilayer has an average
width of  3 nm and the extension of the entire
system along the direction perpendicular to the bilayer 
is of 8.5 nm.

\begin{figure}
 \centering
    \includegraphics[scale=.20]{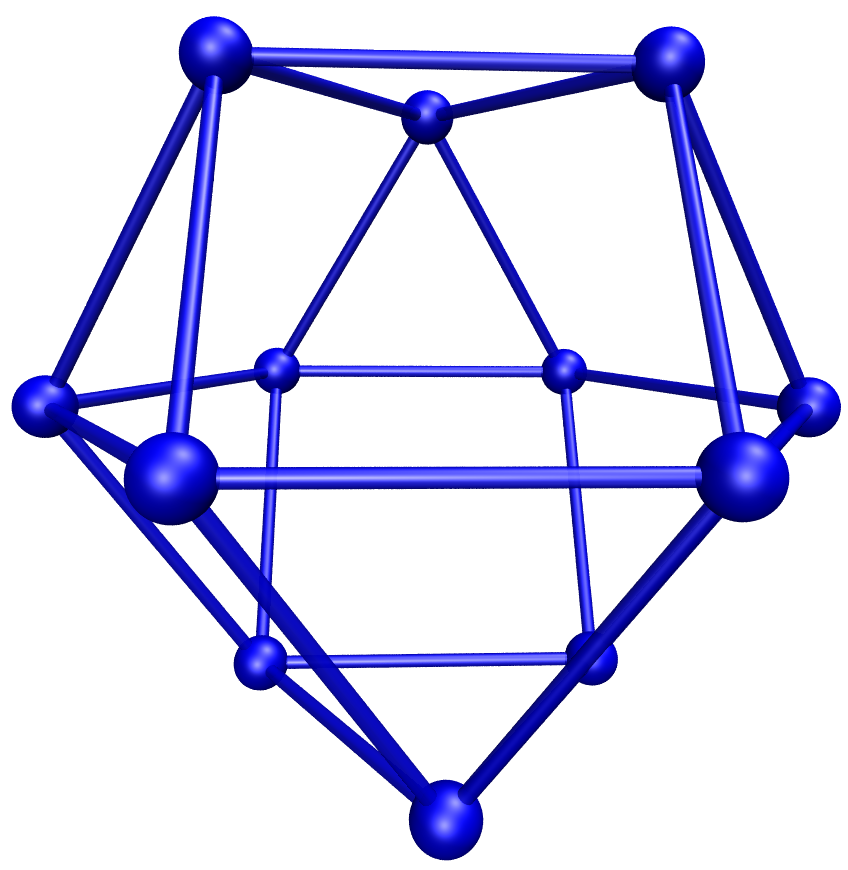}
    \caption{Pictorial representation of the second shell of neighbors in cubic ice (or cuboctahedron, $\bar{C}$). 
    Blue spheres indicate the oxygens positions, while the blue lines emphasize the geometrical structure.}
 \label{fig:Fig0}
\end{figure}

We use the simulation
package NAMD 2.9~\cite{Phillips_JCompChem2005} at a temperature of 303
K and an average pressure of 1 atm. We set  the simulation time step
to 2 fs. We describe the structure of phospholipids and their mutual
interactions  by the recently 
parameterized force field CHARMM36~\cite{Klauda_JPhysChemB2010,Lim_JPhysChemB2012}, 
which is able to reproduce the area per lipid in excellent agreement with experimental data. 
The water model employed in our simulations, consistent with the parametrization of CHARMM36,
is the modified TIP3P~\cite{Jorgensen_JCP1983,MacKerell_JPhysChemB1998}. 
We cut off the Van der Waals interactions at 12 \AA~ with a smooth switching function starting at 10\AA. 
We compute the long ranged electrostatic forces with the help of the particle mesh Ewald 
method~\cite{Essmann_JCP1995}, with a grid space of about 1~\AA. After energy minimization,
we equilibrate the hydrated phospholipid bilayers for 10~ns followed by a production run of 2~ns 
in the $NpT$ ensemble at 1 atm. In the simulations, we control the temperature  by a Langevin 
thermostat~\cite{Berendsen_JPhysChem1984} with a damping coefficient of 0.1~ps$^{-1}$, 
and we control the pressure by a Nos\'{e}-Hoover Langevin barostat~\cite{Feller_JChemPhys1995} 
with a piston oscillation time of 200~fs and a damping time of 100~fs.

\section{Results}\label{results}

\subsection{Structure}

We characterize the local structures adopting as a reference a
cuboctahedron $\bar{C}$ (Fig.~\ref{fig:Fig0}), 
belonging to the class of Archimedean solids enriched with edge-transitivity~\cite{atlas_2016} and describing
the oxygen lattice in the second shell of neighbors in a perfect cubic
I$c$ ice. 
Analogous results can be drawn 
using the anticuboctahedron ($C$) which describes the oxygen lattice in the second shell of neighbors in perfect 
hexagonal I$h$ ice.

%A pictorial representation  of $\bar{C}$ is reported in
%fig.~\ref{fig:Fig0}, where the blue spheres indicate the oxygen
%positions, while the lines connecting the vertices serve to emphasize
%the geometrical structure.
In Fig.~\ref{fig:Fig0} the six-folded ring on the $\sigma$-plane
is a plane of inversion.
In the case of hexagonal ice, one of the two three-folded rings above or below the 
$\sigma$-plane is rotated by $60^{\circ}$.

\begin{figure}
 \centering
    \includegraphics[scale=.33]{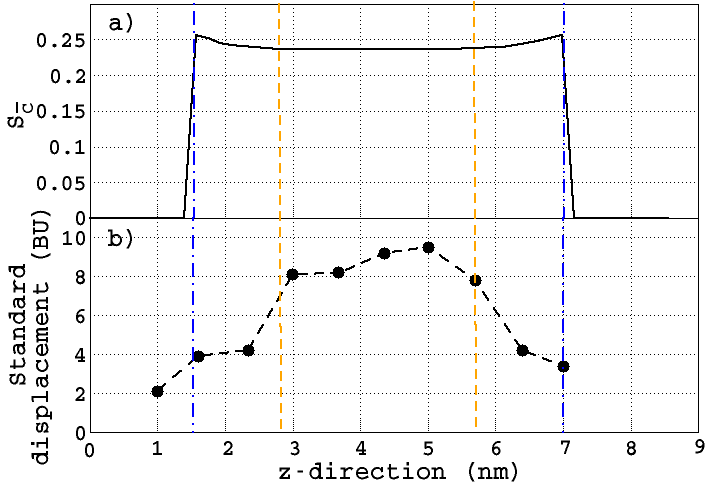}
    \caption{a) Average score $S_{\bar{C}}$ of water molecules
      belonging to a bin centered at 
      distance $z$ from the center of the lipid bilayer and with a bin-width of 1/10 of
      the entire system. The system has periodic boundary conditions
      at $z=8.5$ nm. Each lipid layer has an average extension of 1.5
      nm and the average width of the bilayer is 3 nm.  
    Vertical dashed (orange) lines indicate the region were variations of $S_{\bar{C}}$
    become appreciable.
    Vertical dotted dashed (blue) lines indicate where the water-lipid
    interfaces are on average.
    The sudden drops in $S_{\bar{C}}$,
    at  $z\simeq 1.5$ nm and $z\simeq 7$ nm, delimit the 
    average position of the water-membrane interfaces.
    b) Average {\it standard} displacement of water molecules within
    the bin at distance $z$. For each molecule
    $i$ the displacement is
    normalized with $i$th BU $d_i$.}
 \label{fig:Fig1}
\end{figure}

We divide the entire system along the direction perpendicular to the
bilayer in 10
bins and calculate, for the water molecules within each bin centered at distance 
$z$ from the center of the bilayer, the average score $S_{\bar{C}}$ 
(Fig.~\ref{fig:Fig1}.a).
We find that 
$S_{\bar{C}}$ is approximately constant in a central region 
with 2.8 nm $<z<$ 5.8 nm (vertical dashed orange lines in
Fig.~\ref{fig:Fig1}.a) 
between the lipid layers.
The score
slightly increases moving from the center toward the lipid interfaces 
at $z\lesssim 2.8$ nm and $z\gtrsim 5.8$ nm.
The sudden drops in $S_{\bar{C}}$,
at  $z\simeq 1.5$ nm and $z\simeq 7$ nm, correspond to the average
position of the fluctuating water-membrane interface (vertical dashed-dotted blue lines) 
The behavior of $S_{\bar{C}}$ suggests that the interface has
a significant effect on water structure over a distance $< 1.5$ nm from the interface,
while at larger distance water might be bulk-like, at least for some properties.
%This region, as we will discuss in the following, is {\it bulk-like}
%for some--but not all--of its properties.
%As we will discuss in the
%following, the definition of ``bulk'' is ambiguous and depends upon 
%which characteristics of homogeneous bulk water one is referring to.

\begin{figure}
 \centering
    \includegraphics[scale=.35]{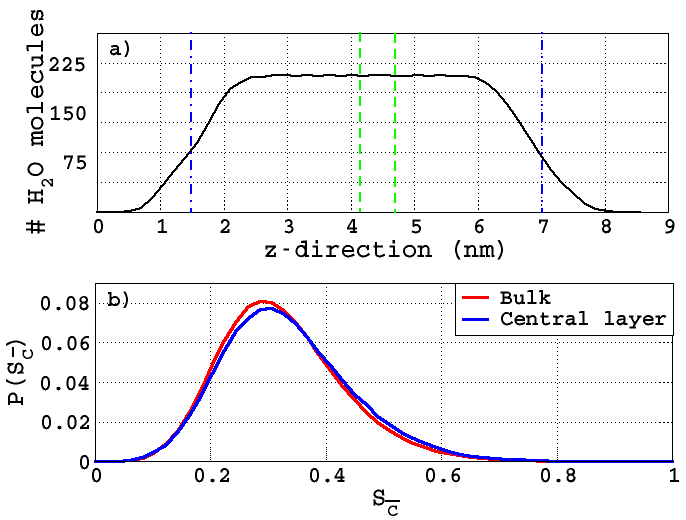}
    \caption{a) Number of water molecules in between the lipid layers
      along the $z$-direction. We perform our structural analysis on
      the molecules within the central layer delimited by the two
      vertical dashed (green) 
    lines at 4 and 4.4 nm.
    b) Distributions of $S_{\bar{C}}^{central}$ (green) and of
    $S_{\bar{C}}^{bulk}$ (red). The degree of order in the IRO in 
    bulk water  is slightly different from that in the central,
    confined,  region.}
 \label{fig:Fig2}
\end{figure}

To better understand how, and  to which extent, confined water becomes bulk-like,
we calculate the number of water molecules as function of the distance
$z$ (or $z'\equiv 8.5-z$) from the center of the membrane
(Fig.~\ref{fig:Fig2}.a). We find that on average water penetrates
within the membrane, that extends up to $z\simeq 1.5$ nm (or $z'\simeq
1.5$ nm, i.e. $z\simeq 7$ nm) and
populates the confined region between the membranes with a density
profile that saturates at $\simeq 1.5$ nm further distance from the
membrane at $z\simeq 3$ nm (or $z'\simeq  3$ nm, i.e. $z=5.5$ nm).

This results are consistent with the analysis of 
a previous studies~\cite{Calero_Mat2016}, where Calero et
al. introduce the intrinsic distance ($\xi$) from the 
membrane by performing a  two-dimensional Voronoi tessellation of the
average plane of the membrane (the $xy$-plane) using the phosphorous and
nitrogen atoms of the phospholipid heads as centers of the Voronoi
cells. The introduction of $\xi\equiv z - z_{\rm Voronoy}$ where
$z_{\rm Voronoy}$ is the $z$ coordinate of the center of the Voronoi
cell, allows one to characterize the water density profile filtering
out the noise induced by the fluctuations of the water-membrane
interface. In particular, the use of $\xi$ emphasizes the penetration
of water within the membrane, the layering of
water near the membrane and the existence of a central region with
approximately constant density profile at $\simeq 0.5$ nm away from the
interface when the hydration level is comparable at that considered
here~\cite{Calero_Mat2016}. Calero et al. show that
the density of water is the same as
bulk under the same thermodynamic conditions at distances $0.5$ nm apart from the membrane.
This observation, therefore, suggests that water in the central region
has almost bulk-like properties.

A further comparison with Calero et al. results~\cite{Calero_Mat2016}
shows  that the increment of $S_{\bar{C}}$ towards the
lipid interface resembles the increase of 
water density $\rho$  along the same directions.
This concomitant increment of $\rho$ and  $S_{\bar{C}}$ might 
seem counterintuitive, however we rationalize it as a consequence of
the isotropic reduction of the distances between all second shell
molecules (not shown). Furthermore, 
the increasing of  $S_{\bar{C}}$, i.e. of
IRO, is also related to the strengthening  of the
interactions of water with the lipid membrane, as 
our analysis of the
hydration water dynamics shows in the following subsection.

Strong of the knowledge from the previous work~\cite{Calero_Mat2016}
and from the present analysis
(Fig.~\ref{fig:Fig2}.a), we focus our structural analysis in a 
central region at a distance of 2.5 nm from the membrane, i.e.,
with $4.0$ nm $<z<4.4$ nm.
We compare the IRO in this region with the IRO of bulk water as
obtained from simulations of $6000$ water molecules  
at the same thermodynamic conditions and with the same interaction potential.
In particular, we compute $S_{\bar{C}}$ only for those water molecules
with all their $12$ second neighbors within the region with
constant density profile ($2.5$ nm$ <z<5.5$ nm). We find that the $S_{\bar{C}}$ for the central region follows a
distribution $P(S_{\bar{C}})$ that differs 
slightly, but neatly, from the bulk case  (Fig.~\ref{fig:Fig2}.b).
In particular, bulk water is slightly less ordered than confined water.
This result suggests that the
lipid membrane affects the IRO of water at distances as far as $\simeq
2.5$ nm from the interface, corresponding to approximately seven water diameters. We infer that 
a possible source of such long range effect are the fluctuations of the membrane, which
create dynamical density heterogeneities.  

\subsection{Dynamics}

The effect of the interface on density and structure 
is accompanied by changes in the
dynamical behavior of confined water.
For example, Calero et al.~\cite{Calero_Mat2016} find that 
 the coefficient of diffusion parallel to the membrane and the
 rotational relaxation time of water confined between lipid membranes 
approach the bulk values within 25\% when water moves further than 0.5
nm from the interface.

Here,  to study the dynamics of the membrane-confined water we collect data 
over $1$ ns trajectories and 
calculate  the {\it standard} displacement, defined as the 
distance traveled by a  
water molecule $i$ normalized to the $i$th O--O
BU $d_i$, averaged over all the molecules $i$. We perform the
calculation  bin by bin (Fig.~\ref{fig:Fig1}.b).
%The standard displacement 
%is indicative of the number of O--O average distances a
%water molecule travels.
%, normalized with respect to the nearest neighbor $O-O$ distance.

Our data show that water molecules in the central region between the lipid layers 
travel up to $\simeq$ 10 BU,  while they significantly slow down 
approaching the interface where  $S_{\bar{C}}$ increases  (Fig.~\ref{fig:Fig1}.b). 
For the water within 1 nm from the interface, i.e.  
at  $z\lesssim 2.5$ nm and $z\gtrsim 6$ nm,
the displacement drops by 60\% to $\simeq 4$ BU.
For water penetrating the membrane, e.g. $z\lesssim 1$ nm,
the translational motion reduces by a further 50\% to 
$\simeq 2$ BU.
A standard displacement  $<1$ BU would correspond to water
molecules rattling in the cage formed by their nearest neighbors. This
case  would represent
a liquid in which the translational degrees of freedom are frozen.
Therefore, our data show a drastic effect of the interface on the
water dynamics at least within a distance of 1 nm from the membrane. 

To better understand how strong this effect, we compare with the
bulk case.
The definition of standard displacement 
depends on the total time of the simulation,
therefore we compare with benchmark cases 
with the same simulation time. As a reference case we use
$1$ ns trajectories for bulk TIP4P/2005-water. 
Although we simulate here TIP3P-water and although its properties are
fairly different from than of TIP4P/2005-water, a qualitative comparison  
of results with the two models is still possible.
In particular, we observe that 
close to the lipid interfaces water reaches values similar to
those found for supercooled TIP4P/2005-water after  $1$ ns
simulation time~\cite{Martelli_2016}.

This suggests that the increment in the IRO discussed in the previous
subsection can be ascribed to the
significant slowing down in the translational degrees of freedom. 
Indeed, a decrease in the
translational diffusion can be interpreted as a reduction in the magnitude of the thermal noise.
Hence,  the configurational entropy contribution to the free energy
minimization reduces its relevance while the potential energy term,
due to the electrostatic interactions, acquires more 
weight leading to a more structured configuration.

Further insights on the effects that dynamical properties have on
structural properties
%as captured by $S_{\bar{C}}$
can be achieved by calculating 
the correlation functions 
\begin{equation}
  C_{\vec{A}}(t)=\left<\vec{A}(t)\cdot\vec{A}(0)\right>
  \label{eq:corr}
\end{equation}
being $\vec{A}$ either (i) the dipole vector $\vec{\mu}$ or 
(ii) the
$\overrightarrow{\rm{OH}}$ vector.
In particular, Calero et al. show~\cite{Calero_Mat2016}
that $C_{\vec{\mu}}(t)$ encompasses very insightful information about dynamical 
processes involving molecular rotations.

\begin{figure}
 \centering
    \includegraphics[scale=.35]{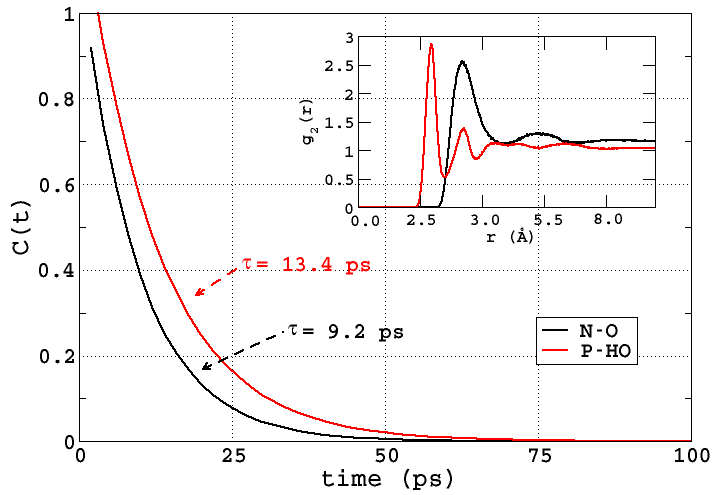}
    \caption{Time correlation functions $C(t)$ for the N--O vector
      (black) and for the P--HO vector (red) at the water-membrane
      interface. Exponential fits of the calculations lead to estimate a correlation
      time that is larger for the P--HO vector. We calculate $C(t)$
      averaging over distances up to the first minimum in the 
 N--O (black) and the P--HO (red) radial distribution
    functions (inset).}
 \label{fig:Fig4}
\end{figure}

On the other hand, DMPC lipids contain N- and P- heads, with 
N interacting mostly with water oxygens and P interacting mostly with
water hydrogens, indicated as HO.
It is reasonable to guess that the N--O and the P--HO interactions
have different strengths. This assumption can be directly tested by
calculating
the time correlation function 
$C(t)\equiv \left<\vec{\delta}(t)\cdot\vec{\delta}(0) \right>$,
where $\vec{\delta}$ is the N--O vector, or the P--HO vector (Fig.~\ref{fig:Fig4}).
We find that the P--HO vector has a longer lifetime compared to that of
the N--O vector, indicating that
the interactions
between P and water H's are stronger than the interactions between N and O. 
This conclusion is also consistent with the observation
that the P--HO radial distribution functions (r.d.f.) has the first peak at
shorter distance than the r.d.f. for N--O (inset in Fig.~\ref{fig:Fig4}).

Based upon the observation that  the N--O and the P--HO interactions
have different strengths, we guess that  rotations around  $\vec{\mu}$  are different from those
around  $\overrightarrow{\rm{OH}}$.
In bulk water at these thermodynamic 
conditions the two rotations are, instead, 
indistinguishable.
Here, we expect them to play a different role in the structure of
water at the interface.

\begin{figure}[!]
 \centering
    \includegraphics[scale=.35]{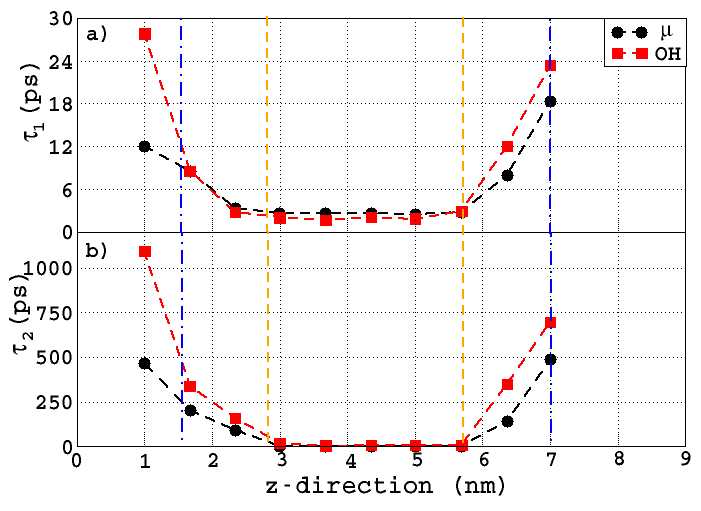}
    \caption{a) Fast relaxation times $\tau_{1}$ for $\vec{\mu}$ (black dots)
      and for OH vector
      %$\vec{\rm{OH}}$
      (red squares) as a function of  $z$.
            Vertical (orange) lines indicate the region where variations 
            of $S_{\bar{C}}$ become appreciable, and are at the same
            $z$ as in Fig.~\ref{fig:Fig1}.
            b) Same as panel a) but for the slow relaxation times $\tau_2$.}
 \label{fig:Fig3}
\end{figure}

We, therefore,  calculate $ C_{\vec{\mu}}(t)$ and $C_{\overrightarrow{\rm{OH}}}(t)$
  for water molecules belonging to bins centered at different $z$
  (not shown) and fit them with a double exponential
  function~\footnote{Calero et al.~\cite{Calero_Mat2016} 
    correctly observe that
    it is formally more correct to integrate the correlation function to get 
    the rotational time
    $\tau^{rot}\equiv\int_0^{\infty}C_{\vec{A}}(t)dt$.
    Although the analysis of  $\tau^{rot}$ gives qualitatively similar
    conclusions of the one present here, 
    we adopt the common biexponential fitting
because it makes more intuitive to reveal 
the effect of electrostatic interactions on the slow relaxation,
which play an important role in 
our analysis.}.
  From the
  fits we
  extract two characteristic times for
  each $z$, one for each exponential function:
  $\tau_1(z)$, associated to fast relaxation modes, and 
  $\tau_2(z)$,  associated to the   slow  relaxation modes
    (Fig.~\ref{fig:Fig3}).
 We interpret the two families of  
  relaxation processes as a consequence of the 
  dynamical heterogeneities induced by the interaction of water with 
    the lipid layers.

%In panel a) of Fig.~\ref{fig:Fig3} we report the fast relaxation time $\tau_1$, while in panel b) we report 
%the slow relaxation time $\tau_2$ in ps for the $\vec{\mu}$ (black) and for the $\overrightarrow{\rm{OH}}$ (red) vectors,
 %respectively.
We find that water rotations slow down approaching the membrane. In
particular, the distances at which the change is appreciable are the
same as for the translations,
at $z\lesssim 2.8$ nm and $z\gtrsim 5.8$ nm.

We observe that 
approaching the lipid membrane both $\tau_1$ and $\tau_2$ increase.
Interestingly, the relaxation times for 
${\overrightarrow{OH}}$  increase at a pace higher than the relaxation times for 
$\vec{\mu}$.
This observation is consistent with our finding that the P--HO
interaction is stronger than the N--O interaction and can be 
rationalized observing that the lipids carry
different (delocalized) charges on the 
N-heads and on the P-functional groups and that these charges can
affect in different way the rotations around the two vectors.

We find that the rotational relaxation times of water penetrating the
membrane, at $z< 1.5$ nm, largely increase, as it would be expected
for degrees of freedom near to their freezing point. Also in this
case the slowing down is stronger near the P-groups than
near the N-heads.

The slowing down of the rotational dynamics decreases as water moves
from the interface toward 
the central region, with 2.8 nm $\lesssim z\lesssim$ 5.8 nm.
At these distances we find that
 all the relaxation times are almost indistinguishable, with 
$\tau_{1}^{\mu}\simeq\tau_{2}^{\mu}\simeq \tau_{1}^{OH}\simeq\tau_{2}^{OH}$, 
as one would expect in bulk water.

Therefore, the water recovers the bulk dynamic behavior at $\simeq
1.3$ nm away from the membrane interface, corresponding 
approximately to four water diameters.
However, this result seems
to be at variance with our structural analysis, in the previous
section, showing that  the IRO structure is affected as far as
$\simeq 2.5$ nm away from the interface. We understand this
difference as a consequence of the fact that the structural
parameter $S_{\bar{C}}$ has a  higher sensitivity, with respect to
the correlation functions, to small effects due to the
interface. In particular, since we adopt a definition
on the second water hydration shell, the effect is observed on 
$S_{\bar{C}}$ up to three water diameters further than on the
correlation times, reaching a total of seven water diameters,
consistent with our  conclusion of the previous subsection. 

\section{Conclusions}\label{conclusions}

We employ the sensitive order metric $S$ to characterize the structural properties of water confined 
between phospholipid membranes.
Commonly inspected quantities such as, e.g., O-O radial distribution function, molecular displacement and rotational 
correlation functions, do not capture differences between bulk water and the farthest water layer from the lipid surfaces.
Instead, we show that the high sensitivity of $S$ is able to detect small, but not negligible, structural differences at the 
IRO. The lipid membrane, therefore, perturb the structure at the IRO of water as far as $\simeq 2.5$ nm from the lipid 
surfaces. It is worthy to note that this value is limited to the finite size of our sample. Therefore, larger 
simulations could better estimate the extent of such long-range effect.  

We find that the IRO of water increases approaching the membrane
surfaces and is concurrent with the slow down of the translational and rotational 
water dynamics.
Our observations indicate that water translational and rotational correlation times show a 
remarkable increase approaching the membrane.
The translational times
reach values comparable to that of water at supercooled conditions. 
On the other hand, we show that the membrane acts 
unevenly on the rotational degrees of freedom, slowing down the
rotations related to the $\overrightarrow{\rm{OH}}$ vectors more than
the  rotations related to the water dipole vector.
Our calculations reveal that
this is due to the stronger interaction between 
P- groups and water's hydrogen with respect to the interactions
between N- heads and oxygens.

%As a consequence, the $\overrightarrow{\rm{OH}}$ vector decays slower compared to the $\vec{\mu}$ vectors
%and the relative diffusivity is slightly slower for molecules interacting with P- groups than 
%molecules interacting with N- heads. 
Such effect has physical implications on the structure and local density of water: molecules 
interacting with P- groups belong to environments slightly more ordered than molecules interacting 
with N- heads. 
Moreover, water molecules interacting with P- groups have closer second
neighbor than those interacting with  N- heads.
Although only qualitative, these observations are
in agreement with our considerations linking the slowing down of the translational degrees of freedom with the 
increment of $S_{\bar{C}}$.
Further analysis adopting the intrinsic distance $\xi$ from the
membrane~\cite{Calero_Mat2016} and higher statistic will be necessary to make these
observations quantitative.
%The faster decay of the dipole correlation function is furthermore enhanced by the fact that each water molecule has 
%two $\overrightarrow{\rm{OH}}$ vectors, therefore increasing the probability for a P group to be bound with one 
%of the two water hydrogens. \par
%

In conclusions, our results suggest that the effect of the
phospholipid membrane on the structural properties of water extend
as far as $\simeq 2.5$ nm from the membrane,
a larger distance than previously calculated with less sensitive observables.
Bulk density and dynamical properties are recovered at a much shorter distance from the membrane, i.e., 
at $\sim 1$ nm,
showing that the definition of a bulk-like region depends
intrinsically on the property we use to compare with bulk. 
This effect should, therefore, be taken into account when studying properties of water 
at interfaces. Accordingly to our findings, the order 
metric $S$ is an ideal tool to inspect such structural changes.

\acknowledgments{C.C.B. and G.F. thank for financial support the Spanish
Ministry of Economy and Knowledge (MINECO) and the European Fund for
Regional Development (FEDER) with grant  FIS2015-66879-C2-2-P 
and the Barcelona Supercomputing Centre (projects QCM-2014-3-0029
and QCM-2015-3-0023).}

\bibliography{References.bib}

\end{document}